# On-chip electro-optically tunable narrow linewidth Brillouin microlasers implemented in thin film lithium niobate

Chuntao Li[1,11], Jiale Deng[2,11], Xingzhao Huang[2], Xiaochao Luo[3], Renhong Gao[4], Huakang Yu[2,§], Jianglin Guan[1], Jacob B. Khurgin[6,◊], Zhiyuan Li[2], Jintian Lin[3,5,*], and Ya Cheng[1,3,4,7,8,9,10,†]

[1]*State Key Laboratory of Precision Spectroscopy, East China Normal University, Shanghai 200062, China*

[2]*School of Physics and Optoelectronics, State Key Laboratory of Luminescent Materials and Devices, South China University of Technology, Guangzhou 510460, China*

[3]*State Key Laboratory of Ultra-intense laser Science and Technology, Shanghai Institute of Optics and Fine Mechanics, Chinese Academy of Sciences, Shanghai 201800, China*

[4]*The Extreme Optoelectromechanics Laboratory (XXL), School of Physics and Electronic Science, East China Normal University, Shanghai 200241, China*

[5]*University of Chinese Academy of Sciences, Beijing 100049, China.*

[6]*Department of Electrical and Computer Engineering, Johns Hopkins University, Baltimore, MD 21218, USA*

[7]*Shanghai Research Center for Quantum Sciences, Shanghai 201315, China*

[8]*Hefei National Laboratory, Hefei 230088, China*

[9]*Collaborative Innovation Center of Extreme Optics, Shanxi University, Taiyuan 030006, China*

[10]*Collaborative Innovation Center of Light Manipulations and Applications, Shandong Normal University, Jinan 250358, China*

[11]*These authors contributed equally to this work.*

[§]*hkyu@scut.edu.cn*

◊*jakek@jhu.edu*

**jintianlin@siom.ac.cn*

[†]*ya.cheng@siom.ac.cn*

## Abstract

On-chip narrow linewidth microlasers with real-time wavelength tunability are highly desirable for various applications including precision metrology, quantum technology, and coherent information processing. Realizing such laser remains a challenge despite significant advances made by various groups in recent years [Nat. Commun. 13, 5344 (2022); Nature 615, 411 (2023); Appl. Phys. Lett. 124, 131101 (2024); Nat. Photonics 13, 60 (2019)]. In this work, we overcome these hurdles and demonstrate on-chip electro-optically tunable Brillouin microlasers in compact lithium niobate on insulator (LNOI) microdisks with diameters of 31.5 μm and 117.0 μm by using cross-polarized SBS arrangement. A quasi-continuum band of bound shear mechanical modes inside the suspended microdisk are revealed for the first time, allowing feasible phase matching of stimulated Brillouin lasing (SBL). We achieve efficient cross-polarized optomechanical coupling and SBL via the significant photoelastic tensors of lithium niobate (e.g., $p_{41}$=-1.51). This approach yields a 118 Hz intrinsic linewidth and a comparatively low threshold power of 3.15 mW. A real-time electro-optic tuning of the cross-polarized Brillouin scheme with a tuning efficiency of ~93.1 kHz/V is also achieved, further showcasing potential of LNOI platform for next-generation tunable photonic systems.

Narrow linewidth laser sources with a fast wavelength tunability provide spectral purity required for a wide range of applications ranging from precision metrology, quantum technology, to coherent information processing [1-6]. Among other methods, stimulated Brillouin lasers (SBLs) stand out as nonlinear coherent optomechanical interaction between photons and phonons in optical resonators, simultaneously enabling linewidth narrowing of the pump laser by orders of magnitude and flexible wavelength tunning [7-11]. With its narrow megahertz gain linewidth, microwave-compatible gigahertz frequency shift and unique acoustic dissipation mechanism, SBL provides high spectral purity required for a wide range of applications ranging from coherent communications [7-9], next-generation data center networks [12,13], to high precision metrology [14-16], and miniaturized atomic clocks [17]. In recent years, there has been a growing interest in transitioning these highly coherent light sources from laboratory benchtop setups to miniature resonators [18-21], and ultimately extending to photonic integrated platforms [22-30]. Current photonic platforms for on-chip SBLs are focused on isotropic materials such as Si, SiN, $SiO_2$ and chalcogenide, enabling demonstrations of high spectral purity on-chip SBLs and Kerr-Brillouin microcombs [22-30]. Generally, since the Brillouin frequency shift is orders of magnitude lower than optical frequencies and the gain bandwidth is only on the order of magnitude of 100 MHz, on-chip narrow-linewidth operation has been mainly limited to millimeter-to-centimeter-scale resonators (where free spectral ranges match the Brillouin shift) [8,22-24,26-29]. Unfortunately, such demonstrations simultaneously impede intra-cavity electro-optic tuning of the lasers, hinders the excitation of cross-polarized stimulated Brillouin scattering (SBS) in on-chip resonators, and suffer from large modal volumes which is not conducive to both enhancing optomechanical coupling [30] and low pump power operation.

In view of the outlined restrains, the Brillouin gain medium has begun to expand to anisotropic materials such as lithium niobate. For example, cross-polarized Brillouin lasers have emerged as a breakthrough in benchtop systems [31,32], which leverages the non-diagonal photoelastic coefficient ($p_{41}$=-0.151) of lithium niobate for achieving strong cross-polarized SBS gain, enabling applications such as SBS-based polarization conversion, filtering, and thermally stable frequency comb generation. However, the design and fabrication of such SBLs in anisotropic photonic integrated platforms remains a significant challenge due to the difficulties in simultaneous confinement of optical and mechanical modes, dispersion engineering of on-chip microresonators for energy and momentum conservation. Inside the lithium niobate on insulator (LNOI) waveguide, it was found that very few (only two) mechanical modes, that were not coupled to the substrate, provide substantial Brillouin gain, thereby rendering SBS irrelevant to the formation of SBLs [33]. Moreover, the key functionality such as real-time intra-cavity electro-optical tunability of SBLs, a highly desirable feature for many practical photonic systems, is yet to be developed.

In this work, we address these challenges by demonstrating a cross-polarized Brillouin microlaser in compact LNOI microdisks through by employing novel dispersion engineering scheme. As shown in Fig.1(a), the periphery of the resonator is suspended in air, indicating that tight confinements of both optical and acoustic modes can be achieved [34-36]. A quasi-continuum band of bound shear mechanical modes inside the suspended resonator are revealed and allow feasible phase matching of SBL. By exploiting fundamental cavity modes of orthogonal polarizations, we achieve efficient cross-polarized optomechanical coupling and SBL via the large photoelastic tensors [37]. Furthermore, this configuration allows electro-optic tuning of the cross-polarized SBL was demonstrated, representing the smallest on-chip SBL microresonator (31.5-μm diameter) so far [30].

## Results

### Enabling cross-polarized Brillouin lasing in a small diameter LNOI microresonator

Experimentally, we fabricated suspended LNOI microdisk by femtosecond laser photolithography assisted chemo-mechanical etching [38], as depicted in the Figs. 1(a) and (c). For operating at the telecom wavelength of 1550 nm and acoustic velocity of 3600 m/s, acoustic mode with frequency of ~ 10 GHz is requisite to satisfy energy and momentum conversation for allowing efficient SBS. Unlike the situation in waveguide structures [34, 39-41], fruitful shear mechanical modes inside the microresonator are expected due to the suspended periphery, which prohibits the severe acoustic dissipation to the substrate [34-36]. As shown in Supplemental Materials, numerical simulations by commercial finite-element methods revealed the existence of quasi-continuum band of bound shear acoustic modes (predominantly whispering gallery-like modes (w-modes) or dilatational modes (d-modes) and their hybridizations [39]) within the 10 GHz frequency range, which readily facilitated the phase-matching conditions for SBS.

To facilitate the cross-polarized Brillouin microlaser, dual-resonance condition of the cross-polarized optical fundamental modes was proposed. We utilized two sets of fundamental whispering gallery mode (WGM) modes with orthogonal polarizations: $TM_{00}$ and $TE_{00}$ to satisfy the phase-matching condition, as schematically illustrated in Figs. 1(b), (d) and (e). This choice was motivated by the presence of both the significant diagonal and non-diagonal photoelastic tensor elements ($p_{44}$ and $p_{41}$, see Supplemental Materials), which simultaneously facilitates efficient optomechanical coupling between two orthogonally polarized optical modes of similar spatial distributions interacting with one shear acoustic mode (whose mechanical displacement $|u|$ is shown in Fig. 1(f)), and enable fast electro-optic tunability.

Our fabricated LNOI microdisk exhibits a measured diameter of 117 μm, a thickness of 800 nm, and a wedge angle of the sidewall of 15°. Correspondingly, the free spectral ranges (FSRs) are 390 GHz for the $TE_{00}$ mode and 370 GHz for the $TM_{00}$ mode. As the FSRs of the $TM_{00}$ and $TE_{00}$ modes differ by 5 %, dual-resonance condition of the pump of the $TM_{00}$ mode and Stokes light of the $TE_{00}$ mode become accessible every 60 nm over the wavelength tuning range (from 1510 nm to 1610 nm) [26].

To realize the dual-resonance condition, a continuous-wave (CW) pump light at the wavelength of 1555.81 nm was injected into the LNOI microresonator (see Supplemental Materials for more details). A tapered optical fiber with waist of 2 μm was placed in contact with the top surface of the microdisk with a coupled position of 2.7 μm far from the edge of the microdisk (see Supplemental Materials for more details). While increasing the input power of pump light, a backward Stokes Brillouin laser (SBL) signal could be observed at the wavelength of 1555.89 nm recorded by an optical spectral analyzer (OSA), as shown in Fig. 2(a). From the spectrum, we found that the wavelength interval between the pump wavelength and the SBL signal was ~ 0.08 nm, corresponding to Brillouin frequency shift $\Omega_B/2\pi$ of ~10 GHz. This Brillouin frequency shift, equal to the radio-frequency (RF) beat note signal of the pump light and the backward SBL, was further identified at a frequency centered at 10.3 GHz using a high-speed photodetector (PD) connected with a real-time electrical spectral analyzer (ESA), as shown in Fig. 2(b). And the intrinsic linewidth of the detected RF beat note signal was measured to be 3 Hz (see Supplemental Materials for details), which is narrower than 9 Hz previously reported in LNOI platform [43]. The dependence of the backward SBL power on the on-chip pump power was also measured, as shown in Fig. 2(c). Above the pump threshold of 3.15 mW, the SBL power grows linearly with the increasing pump power, a telltale sign of stimulated emission. From linear fitting, the slope efficiency was determined as high as 6.8% with maximized output SBL power of ~0.4 mW. According to the theory of SBL threshold [18,38,42] as in the Supplemental Materials, for pump light in the experiment, we could obtain Brillouin gain $G_{B,exp} = 0.66(W \cdot m)^{-1}$ for pump light of $\lambda$ = 1555.8 nm and cavity circumference of $L = \pi \cdot$ 117 μm. Furthermore, the abundant acoustic modes from the quasi-continuum band facilitate repeated SBL outputs. As demonstrated in Fig. 2(d), by adjusting the pump laser wavelength from 1555.808 nm to 1546.30 nm, the SBL wavelength shifted from 1555.89 nm to 1546.39 nm, corresponding to a spectral range exceeding 9 nm across multiple discrete tuning instances. Based on the experimental results in Fig. 2(a), we calculated optical dispersion and mechanical dispersion as shown in Figs. 2(e) and (f). Corresponding SBL frequencies in Fig. 2(d) are identified to be all around 10 GHz, which lies within quasi-continuum band of acoustic modes. And because of the emergence of the quasi-continuum band of acoustic modes in suspended LNOI microdisks, the cross-polarized SBL has also been demonstrated in the LNOI microdisk with a diameter of only 31.5 μm (see Supplemental Materials for details), representing the smallest on-chip microresonator for SBL [30].

Numerical simulations further confirmed the strong optomechanical coupling between these cross-polarized optical fundamental modes and shear acoustic mode, with predicted highest Brillouin gain exceeding $100(W \cdot m)^{-1}$ (see Supplemental Materials). According to the experimental parameters, the specific optical and mechanical modes involved in the SBL generation can be identified according to the energy and momentum conservation. It is shown that the phase matching of SBS occurred between a $TM_{00}$ (with azimuth mode number of $m_p$ = 422 @ 1555.808 nm) and $TE_{00}$ ($m_s$ = -455 @ 1555.891 nm) with frequency discrepancy of 10.3 GHz. Following the phase matching condition, ($M = m_p - m_s$), we could identify a hybridized shear mode at 10.326 GHz (M = 877), with its mechanical displacement close to the cavity periphery (see Supplemental Materials for more details). As for the identified optical and mechanical modes, further calculation suggested Brillouin gain $G_{B,SIM} = 46.67(W \cdot m)^{-1}$, which is larger than the experimental value. Such discrepancy arises from several reasons. First, we didn't take the anisotropy of lithium niobate into consideration in our simulation, while Brillouin gain is significantly dependent on the varying azimuthal angle $\phi$ [33,43]. Second, the Brillouin gain calculation assumes pump mode and Stokes mode are both perfectly on resonance, which may not be the real case, leading to insufficient nonlinear SBS experimentally. Third, surface roughness and material absorption could play a role in increasing threshold power. Fourth, the tapered fiber coupling condition also plays an important role in determining the SBL threshold, which is not optimized here [27,44].

**Real-time electro-optic (EO) tuning of the Brillouin laser**

The electro-optic tuning capability is one of the most appealing features for lithium niobate (LN), and is highly in demand for real-time tunable SBLs. For the SBL with cross-polarized configuration here, one advantage is that one can use the in-plane (x-y-plane) electric field to modify the refractive index of TE-polarized mode while not affecting the TM-polarized mode. Experimentally, two in-plane chromium (Cr) microelectrodes perpendicular to X principal axe of LN were designed and monolithically integrated on the left and right sides of the microresonator, as shown in the inset of Fig. 3(b). The gap of the electrode pair was set as 139 μm. This electrode pair would then produce electric field oriented along the X principal axe, as shown in Fig. 3(c), and enable the utilization of EO coefficient $r_{22}$, leading to a dominant modification in the in-plane (X-Y plane) refractive index ellipsoid of the half microdisk [45]. Such configuration would ensure electro-optical tuning of effective refractive index of the Stokes $TE_{00}$ optical mode in the microdisk, while not affecting the pump $TM_{00}$ optical mode. Consequently, when the pump wavelength was fixed at 1555.808 nm, the SBL signal exhibited a red-shift in response to the applied bias voltage, as evidenced by the corresponding RF beat note measurements presented in Fig. 3(a). When the bias voltage was increased from 0 V to 200 V, the RF beat note signal shifted with 8.64 MHz, corresponding to an EO tuning efficiency of the SBL of 43.2 kHz/V. This value is smaller than theoretical prediction of 137 kHz/V (see Supplemental Materials). To further enhance the EO tuning efficiency, the distance of the microelectrodes should be reduced. For example, EO tuning was performed in the aforementioned 31.5-μm diameter microdisk, showing a significantly enhanced efficiency to 93.1 kHz/V (see Supplemental Materials), due to the reduced distance between in-plane microelectrodes of 52.0 μm. To be mentioned, such EO tuning could serve as solid evidence that two fundamental cavity modes with orthogonal polarizations are involved in the SBS process. Aside from the EO tunability, the inherent differences between TE and TM polarizations provide a versatile platform for mode control, for example thermo-optic tunability [31] in future.

**Measurement of the lasing linewidth**

The linewidth of the SBL is one of the crucial features to be determined. Since the loaded Q factors of the pump and Stokes optical modes were measured to be $1.21\times10^6$ and $4.61\times10^6$ (see Supplemental Materials), the linewidth of the SBL should be narrow. The phase noise and intrinsic linewidth of the SBL measured by a home-built delayed self-heterodyne interferometer method based on a short-delayed self-heterodyne method [27,29,43], are presented in Supplemental Materials. These measurements revealed an underestimated intrinsic linewidth of approximately 0.5 Hz. To precisely determine the intrinsic linewidth of the SBL, the frequency noise of the backward SBL measured by the commercially available laser phase noise analyzer based on correlated self-heterodyne method [30,46] is shown in Fig. 4(a). The white-frequency-noise floor $N_{wfn}$ of the backward SBL was measured to be 37.4 $Hz^2/Hz$, revealing intrinsic linewidth $L_{st}$ of 117.5 Hz ($L_{st} = N_{wfn} \times \pi$) [47] which is narrower than 280 Hz previously reported in LNOI platform [43]. And the dependence of the intrinsic linewidth of the SBL on the SBL power was also characterized, as the results plotted in Fig. 4(b). The linewidth almost linearly increases as the growth of the SBL power, confirming that the linewidth corresponds to intrinsic linewidth. Such narrow linewidth is resulted from the narrow Brillouin gain bandwidth and the short lifetime of the Brillouin phonon [23,26].

Theoretically, the linewidth of SBL can be given by [39],

$$\Delta v_{SBL} = \frac{\Gamma^2}{(\gamma+\Gamma)^2}\frac{\hbar\omega_s^3}{4\pi Q_T Q_{ex} P_{SBL}} n_{th} + \frac{\gamma^2}{(\gamma+\Gamma)^2}\Delta v_{pump} \tag{1}$$

where the first term represents the fundamental linewidth and second part represents the contributing factor of linewidth from the transferred pump noise. If the linewidth of pump laser is relatively large or satisfying the condition of mechanical lasing [39] (i.e., $\gamma_s \gg \Gamma_m$), the linewidth of SBL was mainly contributed by the pump noise. If Stokes mode dissipation $\gamma_s$ is far smaller than mechanical mode dissipation $\Gamma_m$ ($\gamma_s \ll \Gamma_m$), Stokes field would experience linewidth-narrowing [26]. We can calculate fundamental linewidth to be 53 Hz based on the result from our experiment (see Supplemental Materials) [41,48,49], which agrees well with the measured result. We can see the main contribution of linewidth was coming from the transferred pump laser noise, leading to much larger linewidth than the sub-hertz linewidth as in Ref. [23], which is reasonable since the Pound–Drever–Hall (PDH) technique was not applied here for stabilizing the frequency of the pump laser. Such narrow linewidth was benefited from the cross-polarized SBL configuration, featuring the broad gain bandwidth and high noise suppression factor [31]. Further improvement of the optical Q factor (e.g., one order of magnitude augment) of the sub-millimeter-diameter microresonator together with minimized transferred pump laser noise would enable narrower SBL laser linewidth even to sub-hertz level akin to the much large resonator counterparts [22,23].

## Discussion and Conclusion

A discussion on how to improve the device performances such as threshold [18,39,50], integration of pump laser and SBL [51-55], intrinsic linewidth by leveraging the coherent cancellation of thermo-refractive noise via the pyroelectric-electro-optic effect as demonstrated in Ref. [53], and EO tuning efficiency [56], can be found in the Supplemental Materials. In conclusion, we have demonstrated on-chip narrow-linewidth cross-polarized Brillouin lasers with real time tunability in small LNOI microdisks with diameters down to 31.5 μm, for the first time to our best knowledge.

Note added—During the preparation of the preprint of this manuscript [57], we became aware of the work by K. Ye et al. [43] also reporting SBL in centimeter-scale

LNOI resonators with the same polarization cavity modes. In additional, the frequency doubling of on-chip SBLs has also demonstrated by our group [58].

We thank Peking university Yangtze delta institute of optoelectronics for providing the laser phase noise measurement system (iFN5000).

**References:**

[1] H. Al-Taiy, N. Wenzel, S. Preußler, J. Klinger, and T. Schneider, Ultra-narrow linewidth, stable and tunable laser source for optical communication systems and spectroscopy. Opt. Lett. **39**, 5826 (2014).

[2] N. G. Pavlov, S. Koptyaev, G. V. Lihachev, A. S. Voloshin, A. S. Gorodnitskiy, M. V. Ryabko, S. V. Polonsky, and M. L. Gorodetsky, Narrow-linewidth lasing and soliton Kerr microcombs with ordinary laser diodes. Nat. Photonics **12**, 694 (2018).

[3] W. Loh, S. Yegnanarayanan, F. O'Donnell, and P. W. Juodawlkis, Ultra-narrow linewidth Brillouin laser with nanokelvin temperature self-referencing. Optica **6**, 152 (2019).

[4] T. Lu, L. Yang, T. Carmon, and B. Min, A narrow-linewidth on-chip toroid Raman laser. IEEE J. Quantum Electron. **47**, 320 (2011).

[5] Z. Bai, Z. Zhao, M. Tian, D. Jin, Y. Pang, S. Li, X. Yan, Y. Wang, and Z. Lu, A comprehensive review on the development and applications of narrow-linewidth lasers. Microwave Opt. Technol. Lett. **64**, 2244 (2022).

[6] M. Li, L. Chang, L. Wu, J. Staffa, J. Ling, U. A. Javid, S. Xue, Y. He, R. Lopez-

Rios, T. J. Morin, H. Wang, B. Shen, S. Zeng, L. Zhu, K. J. Vahala, J. E. Bowers, and Q. Lin, Integrated Pockels laser. Nat. Commun. **13**, 5344 (2022).

[7] A. Kobyakov, M. Sauer, and D. Chowdhury, Stimulated Brillouin scattering in optical fibers. Adv. Opt. Photon. **2**, 1 (2010).

[8] K. Y. Yang, D. Y. Oh, S. H. Lee, Q.-F. Yang, X. Yi, B. Shen, H. Wang, and K. Vahala, Bridging ultrahigh-Q devices and photonic circuits. Nat. Photonics **12**, 297 (2018).

[9] Z. Bai, R. J. Williams, O. Kitzler, S. Sarang, D. J. Spence, Y. Wang, Z. Lu, and R. P. Mildren, Diamond Brillouin laser in the visible, APL Photonics **5**, 031301 (2020).

[10] L. Ren, W. Wang, K. Xu, L. Zhu, J. Wang, L. Shi, and X. Zhang, Stimulated Brillouin scattering in micro/nanophotonic waveguides and resonators, Nanophoton. **14**, 1509 (2025).

[11] E. Garmire, Perspectives on stimulated Brillouin scattering, New J. Phys. **19**, 011003 (2017).

[12] W. Loh, D. Gray, R. Irion, O. May, C. Belanger, J. Plant, P. W. Juodawlkis, and S. Yegnanarayanan, Ultralow noise microwave synthesis via difference frequency division of a Brillouin resonator. Optica **11**, 492 (2024).

[13] B. M. Heffernan, J. Greenberg, T. Hori, T. Tanigawa, and A. Rolland, Brillouin laser-driven terahertz oscillator up to 3 THz with femtosecond-level timing jitter. Nat. Photonics **18**, 1263 (2024).

[14] A. Cygan, D. Lisak, P. Morzyński, M. Bober, M. Zawada, E. Pazderski, and R. Ciuryło, Cavity mode-width spectroscopy with widely tunable ultra narrow laser,

Opt. Express **21**, 29744 (2013).

[15] Y.-H. Lai, M.-G. Suh, Y.-K. Lu, B. Shen, Q.-F. Yang, H. Wang, J. Li, S. H. Lee, K. Y. Yang, and K. Vahala, Earth rotation measured by a chip-scale ring laser gyroscope, Nat. Photonics **14**, 345 (2020).

[16] Y.-H. Lai, Y.-K. Lu, M.-G. Suh, Z. Yuan, and K. Vahala, Observation of the exceptional-point-enhanced Sagnac effect, Nature **576**, 65 (2019).

[17] W. Loh, J. Stuart, D. Reens, C. D. Bruzewicz, D. Braje, J. Chiaverini, P. W. Juodawlkis, J. M. Sage, and R. McConnell, Operation of an optical atomic clock with a Brillouin laser subsystem, Nature **588**, 244 (2020).

[18] I. S. Grudinin, A. B. Matsko, and L. Maleki, Brillouin Lasing with a $CaF_2$ whispering gallery mode resonator, Phys. Rev. Lett. **102**, 043902 (2009).

[19] M. Tomes and T. Carmon, Photonic micro-electromechanical systems vibrating at $X$-band (11-GHz) rates, Phys. Rev. Lett. **102**, 113601 (2009).

[20] S. Zhu, B. Xiao, B. Jiang, L. Shi, and X. Zhang, Tunable Brillouin and Raman microlasers using hybrid microbottle resonators, Nanophotonics **8**, 931 (2019).

[21] H. Zhang, T. Tan, H.-J. Chen, Y. Yu, W. Wang, B. Chang, Y. Liang, Y. Guo, H. Zhou, H. Xia Q. Gong, C. W. Wong, Y. Rao, Y.-F. Xiao, and B. Yao, Soliton microcombs multiplexing using intracavity-stimulated Brillouin lasers, Phys. Rev. Lett. **130**, 153802 (2023).

[22] J. Li, H. Lee, T. Chen, and K. J. Vahala, Characterization of a high coherence, Brillouin microcavity laser on silicon, Opt. Express **20**, 20170 (2012).

[23] S. Gundavarapu, G. M. Brodnik, M. Puckett, T. Huffman, D. Bose, R. Behunin, J.

Wu, T. Qiu, C. Pinho, N. Chauhan, J. Nohava, P. T. Rakich, K. D. Nelson, M. Salit, and D. J. Blumenthal, Sub-hertz fundamental linewidth photonic integrated Brillouin laser, Nat. Photonics **13**, 60 (2019).

[24] I. V. Kabakova, R. Pant, D.-Y. Choi, S. Debbarma, B. Luther-Davies, S. J. Madden, and B. J. Eggleton, Narrow linewidth Brillouin laser based on chalcogenide photonic chip, Opt. Lett. **38**, 3208 (2013).

[25] B. J. Eggleton, C. G. Poulton, P. T. Rakich, M. J. Steel, and G. Bahl, Brillouin integrated photonics, Nat. Photonics **13**, 664 (2019).

[26] N. T. Otterstrom, R. O. Behunin, E. A. Kittlaus, Z. Wang, and P. T. Rakich, A silicon Brillouin laser, Science **360**, 1113 (2018).

[27] Y. Bai, M. Zhang, Q. Shi, S. Ding, Y. Qin, Z. Xie, X. Jiang, and M. Xiao, Brillouin-Kerr soliton frequency combs in an optical microresonator, Phys. Rev. Lett. **126**, 063901 (2021).

[28] M. Zhang, S. Ding, X. Li, K. Pu, S. Lei, M. Xiao, and X. Jiang, Strong interactions between solitons and background light in Brillouin-Kerr microcombs, Nat. Commun. **15**, 1661 (2024).

[29] Y. Li, D. Xia, H. Cheng, L. Luo, L. Wang, S. Zeng, S. Yang, L. Li, B. Chen, B. Zhang, and Z. Li, Low-loss compact chalcogenide microresonators for efficient stimulated Brillouin lasers, Opt. Lett. **49**, 4529 (2024).

[30] M. Wang, Z.-G. Hu, C. Lao, Y. Wang, X. Jin, X. Zhou, Y. Lei, Z. Wang, W. Liu, Q.-F. Yang, and B.-B. Li, Taming Brillouin optomechanics using supermode microresonators, Phys. Rev. X **14**, 011056 (2024).

[31] M. Nie, J. Musgrave, and S.-W. Huang, Cross-polarized stimulated Brillouin scattering-empowered photonics, Nat. Photon. in press (doi:10.1038/s41566-025-01680-7) (2025).

[32] M. Nie, J. Musgrave, K. Jia, J. Bartos, S. Zhu, Z. Xie, and S.-W. Huang, Turnkey photonic flywheel in a microresonator-filtered laser, Nat. Commun. **15**, 55 (2024).

[33] C. C. Rodrigues, N. J. Schilder, R. O. Zurita, L. S. Magalhães, A. Shams-Ansari, F. J. L. dos Santos, O. M. Paiano, T. P. M. Alegre, M. Lončar, and G. S. Wiederhecker, Cross-polarized stimulated Brillouin scattering in lithium niobate waveguides, Phys. Rev. Lett. **134**, 113601 (2025).

[34] C. C. Rodrigues, R. O. Zurita, T. P. M. Alegre, and G. S. Wiederhecker, Stimulated Brillouin scattering by surface acoustic waves in lithium niobate waveguides, J. Opt. Soc. Am. B **40**, D56 (2023).

[35] Y.-H. Yang, J.-Q. Wang, Z.-X. Zhu, X.-B. Xu, Q. Zhang, J. Lu, Y. Zeng, C.-H. Dong, L. Sun, G.-C. Guo, and C.-L. Zou, Stimulated Brillouin interaction between guided phonons and photons in a lithium niobate waveguide, Sci. China Phys. Mech. Astron. **67**, 214221 (2024).

[36] W. Wang, Y. Yu, Y. Li, Z. Bai, G. Wang, K. Li, C. Song, Z. Wang, S. Li, Y. Wang, Z. Lu, Y. Li, T. Liu, and X. Yan, Tailorable Brillouin light scattering in a lithium niobate waveguide, Appl. Sci. **11**, 8390 (2021).

[37] R. Weis and T. Gaylord, Lithium niobate: summary of physical properties and crystal structure. Appl. Phys. A **37**, 191 (1985).

[38] R. Wu, J. Zhang, N. Yao, W. Fang, L. Qiao, Z. Chai, J. Lin, and Y. Cheng, Lithium

niobate micro-disk resonators of quality factors above $10^7$, Opt. Lett. **43**, 4116 (2018).

[39] G. S. Wiederhecker, P. Dainese, and T. P. Mayer Alegre, Brillouin optomechanics in nanophotonic structures, APL Photonics **4**, 071101 (2019).

[40] X.-X. Su, Z.-L. Dou, and H. P. Lee, Stimulated Brillouin scattering in a sub-wavelength anisotropic waveguide with slightly-misaligned material and structural axes: misalignment-sensitive behaviors and underlying physics, J. Opt. **24**, 045002 (2022).

[41] Z. Yuan, H. Wang, L. Wu, M. Gao, and K. Vahala, Linewidth enhancement factor in a microcavity Brillouin laser, Optica **7**, 1150 (2020).

[42] H. Lee, T. Chen, J. Li, K. Y. Yang, S. Jeon, O. Painter, and K. J. Vahala, Chemically etched ultrahigh-Q wedge-resonator on a silicon chip, Nat. Photonics **6**, 369 (2012).

[43] K. Ye, H. Feng, Y. Klaver, A. Keloth, A. Mishra, C. Wang, and D. Marpaung, Surface acoustic wave stimulated Brillouin scattering in thin-film lithium niobate waveguides, arXiv preprint arXiv:2311.14697v2 (2024); K. Ye, H. Feng, R. te Morsche, C. Wei, Y. Klaver, A. Mishra, Z. Zheng, A. Keloth, A. T. Işık, Z. Chen, C. Wang, and D. Marpaung, Integrated Brillouin Photonics in thin-film lithium niobate, Sci. Adv. **11**, eadv4022 (2025).

[44] N. Chauhan, A. Isichenko, K. Liu, J. Wang, Q. Zhao, R. O. Behunin, P. T. Rakich, A. M. Jayich, C. Fertig, C. W. Hoyt, and D. J. Blumenthal, Visible light photonic integrated Brillouin laser, Nat. Commun. **12**, 4685 (2021).

[45] J. E. Toney. *Lithium niobate photonics* (Artech House, 2015).

[46] Z. Yuan, H. Wang, P. Liu, B. Li, B. Shen, M. Gao, L. Chang, W. Jin, A. Feshali, M. Paniccia, J. Bowers, and K. Vahala, Correlated self-heterodyne method for ultra-low-noise laser linewidth measurements, Opt. Express **30**, 25147 (2022).

[47] D. Xu, F. Yang, D. Chen, F. Wei, H. Cai, Z. Fang, and R. Qu, Laser phase and frequency noise measurement by Michelson interferometer composed of a 3×3 optical fiber coupler, Opt. Express **23**, 22386 (2015).

[48] T. J. Kippenberg, S. M. Spillane, B. Min, and K. J. Vahala, Theoretical and experimental study of stimulated and cascaded Raman scattering in ultrahigh-Q optical microcavities, IEEE J. Select. Topics Quantum Electron. **10**, 1219 (2004).

[49] D. Jin, Z. Bai, Z. Zhao, Y. Chen, W. Fan, Y. Wang, R. P. Mildren, and Z. Lv, Linewidth narrowing in free-space running diamond Brillouin lasers, High Power Laser Sci. Eng. **11**, e47 (2023).

[50] O. Florez, P. F. Jarschel, Y. A. Espinel, C. M. B. Cordeiro, T. P. Mayer Alegre, G. S. Wiederhecker, and P. Dainese, Brillouin scattering self-cancellation, Nat. Commun. **7**, 11759 (2016).

[51] Q. Luo, F. Bo, Y. Kong, G. Zhang, and J. Xu, Advances in lithium niobate thin-film lasers and amplifiers: a review, Adv. Photon. **5**, 034002 (2023).

[52] J. Zhou, Y. Zhu, B. Fu, J. Chen, H. Song, Z. Zhang, J. Yu, J. Liu, M. Wang, J. Qi, and Y. Cheng, An integrated electro-optically tunable multi-channel interference cavity laser, Laser Photonics Rev. **19**, 2401664 (2025).

[53] J. Zhang, Z. Li, J Riemensberger, G. Lihachev, G. Huang, and T. J. Kippenberg, Fundamental charge noise in electro-optic photonic integrated circuits, Nat. Phys.

**21**, 304 (2025).

[54] Z. Xie, F. Bo, J. Lin, H. Hu, X. Cai, X.-H. Tian, Z. Fang, J. Chen, M. Wang, F. Chen, Y. Cheng, J. Xu, and S. Zhu, Recent development in integrated lithium niobate photonics, Adv. Phys. X **9**, 2322739 (2024).

[55] M. Li, L. Chang, L. Wu, J. Staffa, J. Ling, U. A. Javid, S. Xue, Y. He, R. Lopez-rios, T. J. Morin, H. Wang, B. Shen, S. Zeng, L. Zhu, K. J. Vahala, J. E. Bowers, and Q. Lin, Integrated Pockels laser, Nat. Commun. **13**, 5344 (2022).

[56] G. Chen, Y. Gao, H.-L. Lin, and A. J. Danner, Compact and efficient thin-film lithium niobate modulators, Adv. Photon. Res. **4**, 2300229 (2023); G. Chen, H.-L. Lin, J. D. Ng, and A. J. Danner, Integrated electro-optic modulator in Z-cut lithium niobate thin film with vertical structure, IEEE Photon. Technol. Lett. **33**, 1285 (2021).

[57] C. Li, J. Deng, X. Huang, X. Luo, R. Gao, J. Lin, H. Yu, J. Guan, Z. Li, and Y. Cheng, Sub-kilohertz intrinsic linewidth stimulated Brillouin laser in integrated lithium niobate microresonators, arXiv preprint arXiv:2411.17443 (2024).

[58] X. Luo, C. Li, X. Huang, J. Lin, R. Gao, Y. Yao, Y. Qiu, Y. Yang, L. Wang, H. Yu, and Y. Cheng, Visible Brillouin-quadratic microlaser in a high-Q thin-film lithium niobate microdisk, arXiv preprint arXiv:2506.08615 (2025).

**Captions of figures:**

FIG. 1 (Color online) Schematic diagram of the stimulated Brillouin lasing. (a) Sketch of the waves in backward Stokes Brillouin laser (SBL) in a suspended LNOI microresonator coupled with a nano-waveguide. (b) Schematic transmission spectra for the one set of fundamental cavity modes with orthogonal polarizations, i.e., $TM_{00}$ (red line) and $TE_{00}$ (black line). Cross-polarized SBS occurs while the dual-resonance condition is satisfied for the pump ($\omega_P$) and Stokes ($\omega_s$) modes. (c) Zoom-in scanning-electron-microscope image of the microresonator, showing a suspended microstructure. The modal profiles of (d) $TM_{00}$ and (e) $TE_{00}$ optical modes, and (f) mechanical displacement $|u|$ of hybridized WGM-like and dilatational shear acoustic modes whose overlap is responsible for strong optomechanical coupling.

FIG. 2 (Color online) Low-threshold SBL. (a) Optical spectrum of the backward SBL signal. (b) The radio-frequency (RF) beat note signal generated by mixing the pump light and the SBL. (c) Power of backward SBL as a function of the on-chip pump power. (d) Optical spectrum of the tuning of the SBL. (e) Modal dispersion of optical modes in the LNOI microdisk. The phase matching of SBS was found to occur between a $TM_{00}$ ($m$ = 422 @ 1555.808 nm) and $TE_{00}$ ($m$ = -455 @ 1555.891 nm) with frequency discrepancy of 10.3 GHz, as depicted by the red arrow. (f) Modal dispersion of shear mechanical modes in the LNOI microdisk. Several discrete acoustic mode families are shown in red, while the gray-shaded region represents a quasi-continuum band formed by densely packed higher-order modes at higher frequencies. To be noted, the boundary of the gray-shaded region in (f) does not correspond to a precisely defined threshold but is instead represented as a schematic illustration. The specific modal dispersion of shear mechanical modes in the LNOI microdisk can be found in the Supplemental Materials (Fig. S1).

FIG. 3 (Color online) Electro-optic tuning of the SBL. (a) RF beat note signal shifts with the varied bias voltage. (b) RF beat note signal frequency drift as a function of the applied bias voltage. (c) Electric potential produced by the bias voltage applied on the microelectrodes.

FIG. 4 (Color online) Measurement of the intrinsic linewidth of the SBL. (a) Frequency noise spectrum of the SBL measured by correlated self-heterodyne method. (b) Intrinsic linewidth of the SBL as a function of the SBL power (blue lines) and the theoretical calculation of the SBL linewidth (red curve).

**Fig. 1**

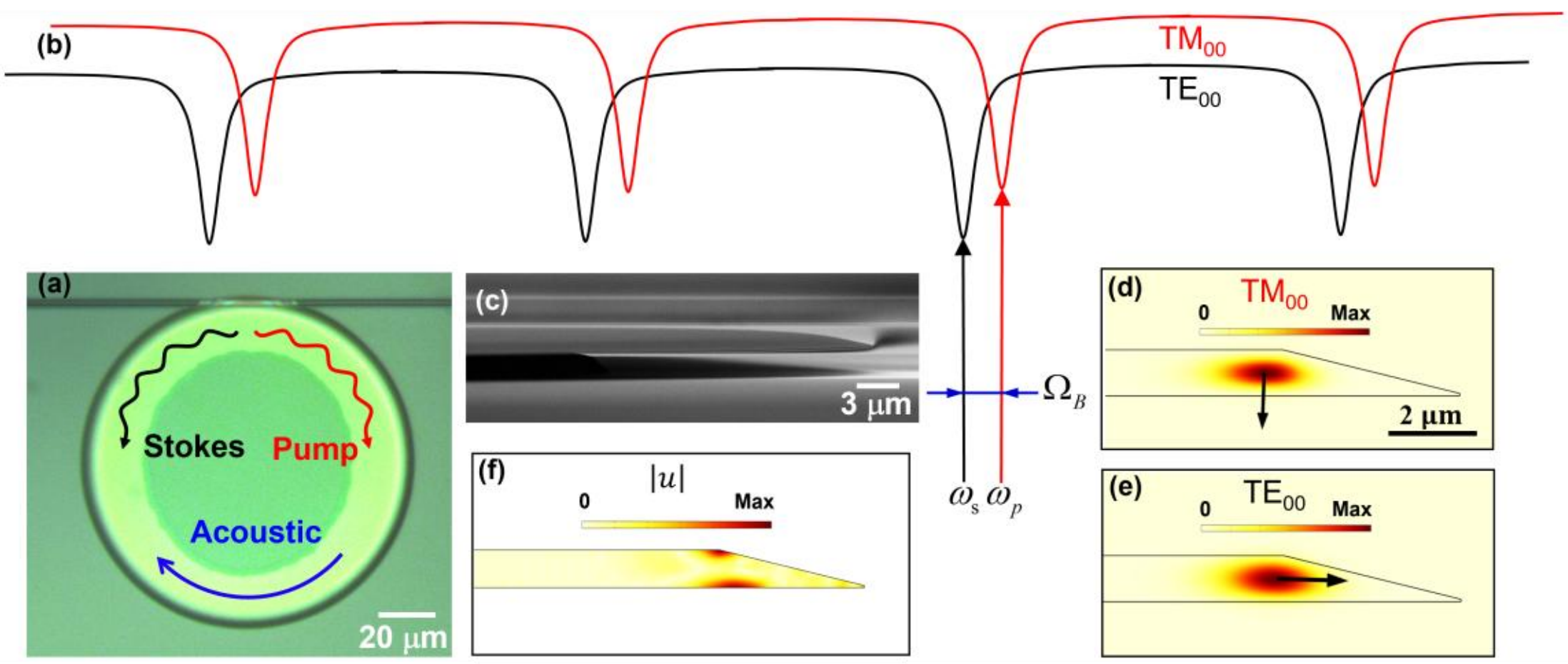

Fig. 2

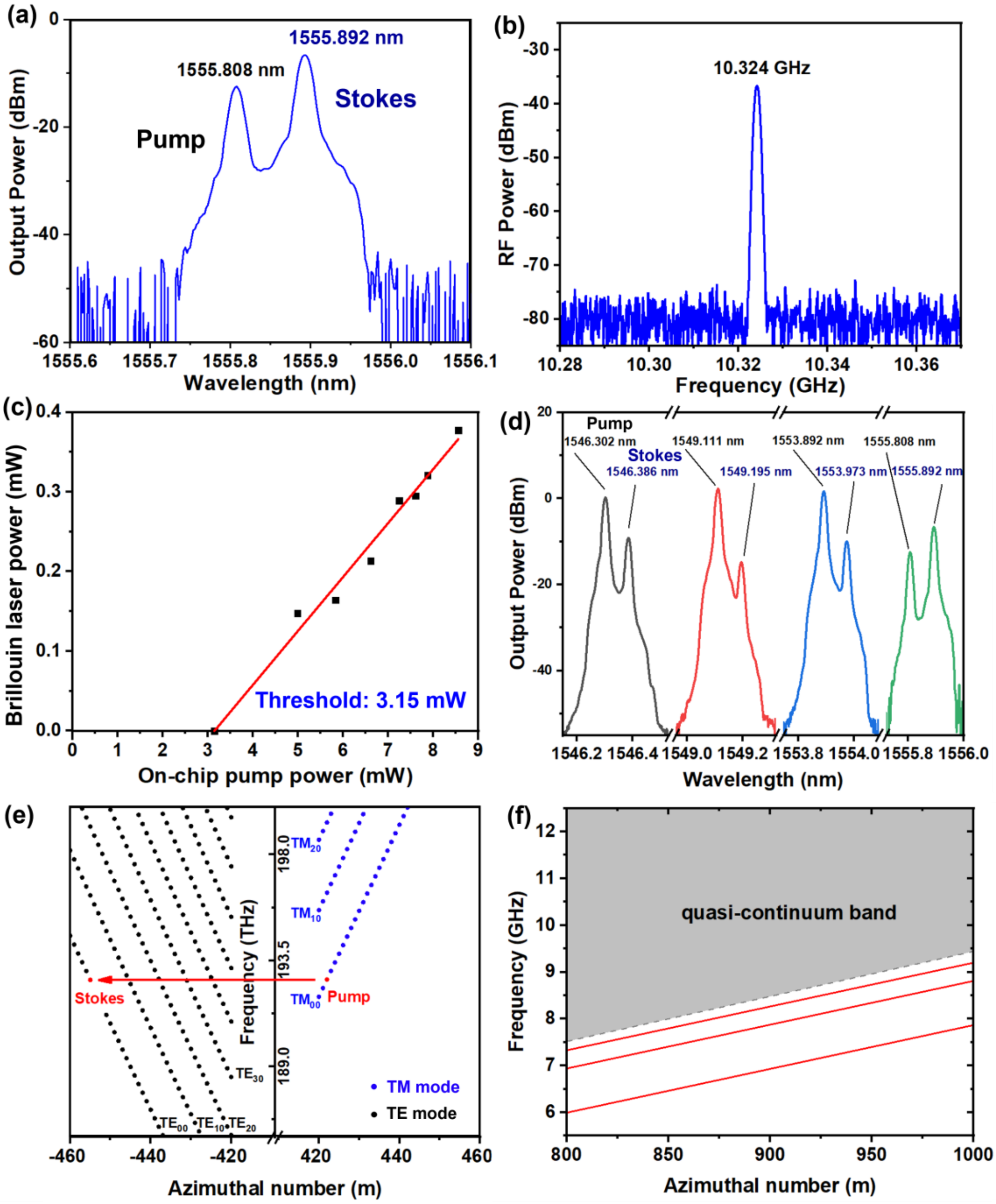

## Fig. 3

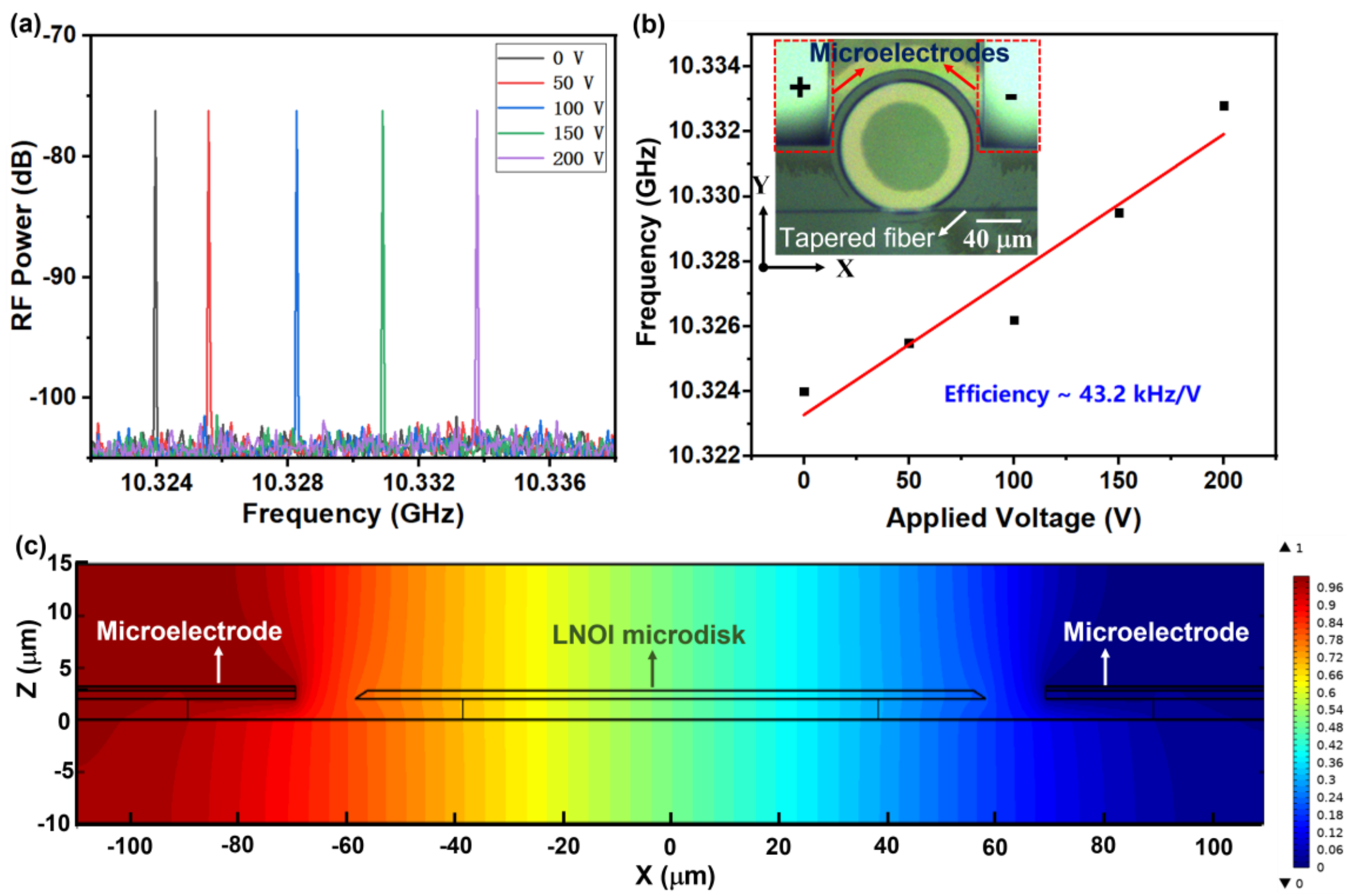

**Fig. 4**

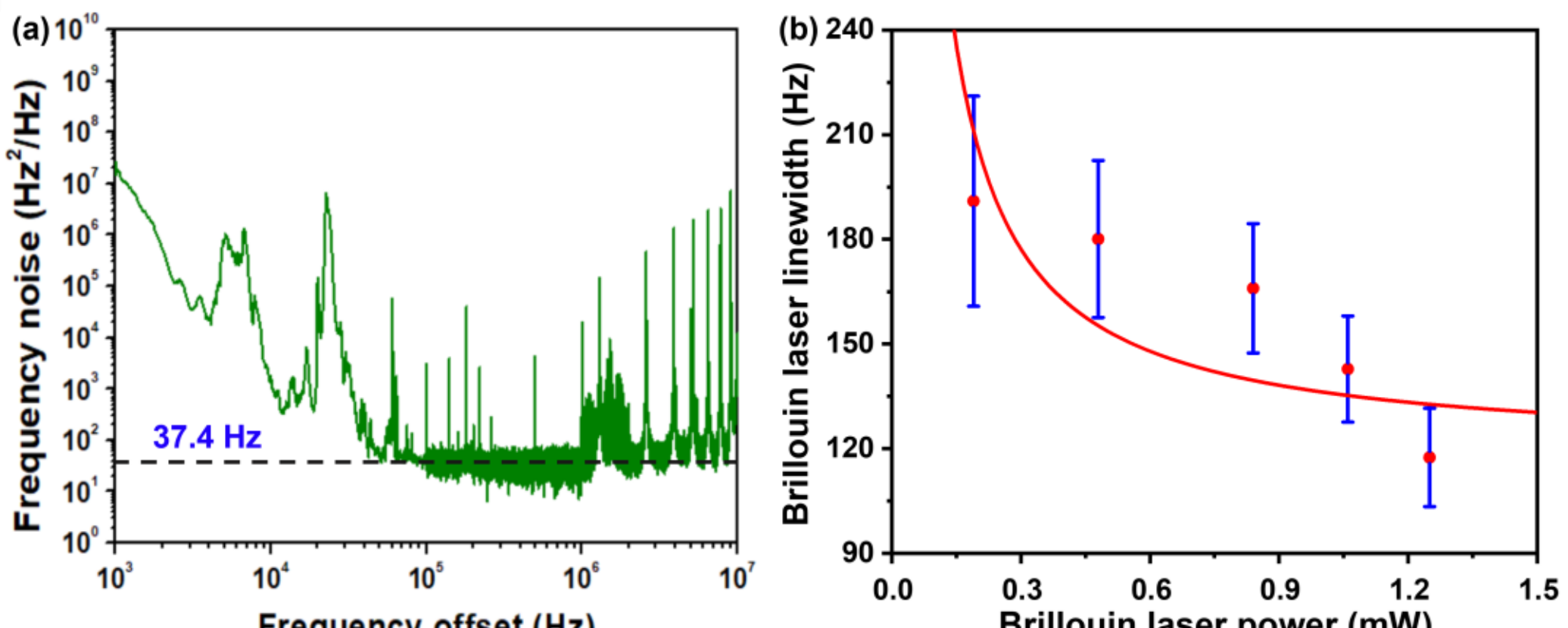